\documentstyle[preprint,aps]{revtex}
\tighten
\begin{document}
\draft
\title{Equivalence of Fokker-Planck approach and non-linear
$\sigma$-model for disordered wires in the unitary symmetry class}

\author{K. Frahm}

\address{Service de Physique de l'\'Etat condens\'e.
	CEA Saclay, 91191 Gif-sur-Yvette, France.}

\date{\today}

\maketitle

\begin{abstract}
The exact solution of the Dorokhov-Mello-Pereyra-Kumar-equation
for quasi one-dimensional disordered
conductors in the unitary symmetry class is employed
to calculate all $m$-point correlation functions by
a generalization of the method of orthogonal polynomials.
We obtain closed expressions for the first two
conductance moments which are valid for the whole range of length scales
from the metallic regime ($L\ll Nl$) to the insulating regime ($L\gg Nl$)
and for arbitrary channel number.
In the limit $N\to\infty$ (with $L/(Nl)=const.$) our expressions agree
exactly with those of the non-linear $\sigma$-model derived from
microscopic Hamiltonians.
\end{abstract}


\narrowtext

The current understanding of transport and localization in
quasi one-dimensional disordered wires
is essentially based on two competing approaches.
The first one \cite{dorokhov,mello2,mello3,mello4,macedo}
is a random matrix approach in which the
transfer matrix is written as a product of statistical independent
building blocks modelling the quasi one-dimensional
structure of the wire.
The distribution of each building block is calculated by maximizing
its entropy (local maximum entropy approach). This procedure relies on
the assumption that all scattering channels are equivalent.
The probability distribution $p(\{\lambda_i\},L)$
(the $\lambda_i$ denote the usual radial coordinates which are
related to the transmission eigenvalues via $T_i=(1+\lambda_i)^{-1}$)
is determined by a Fokker-Planck equation (henceforth called
DMPK-equation) describing the evolution with
the length $L$ of the wire :
\begin{eqnarray}
\nonumber
\frac{\partial p(\{\lambda_i\},L)}{\partial L}&=&
\frac{2}{\xi} \sum_i \frac{\partial}{\partial \lambda_i}
J(\{\lambda_i\})\,\lambda_i(1+\lambda_i)\\
\label{fp_eq}
&&\qquad\qquad\times\frac{\partial}{\partial \lambda_i}
\frac{p(\{\lambda_i\},L)}{J(\{\lambda_i\})}\quad,\\
\label{jj_def}
J(\{\lambda_i\})&=&\prod_{i>j} |\lambda_i-\lambda_j|^\beta\quad.
\end{eqnarray}
The three possible values of $\beta=1,2,4$ correspond to the orthogonal,
unitary or symplectic symmetry classes, respectively, and
$\xi=(\beta N+2-\beta)\,l$ is the localization length.

An alternative description of quasi one-dimensional wires
starts from a more direct, microscopic approach.
The metallic regime ($l\ll L\ll Nl$) is treated in terms of
diagrammatic perturbation theory \cite{altshuler,lee}.
To extend the results to all length scales it is necessary to consider
the one-dimensional supersymmetric non-linear $\sigma$-model
\cite{efetov,zirnbauer1,zirnbauer2,iida}.
Developing the Fourier analysis on the matrix space defining the
$\sigma$-model,
the average of the conductance \cite{zirnbauer1} and its variance
\cite{zirnbauer2} have been calculated for all length scales from the
metallic to the insulating regime ($L\gg Nl$).
Since the one-dimensional $\sigma$-model can be derived from
a Hamiltonian with white-noise potential \cite{efetov},
Wegner's $N$-orbital model
\cite{iida}, and a random band matrix approach for the Hamiltonian
\cite{fyodorov}, it provides a description of quasi one-dimensional
disordered conductors that is rather independent of microscopic details.

The results for the universal conductance fluctuations and
weak-localization corrections in the metallic regime obtained
by the Fokker-Planck approach and by the $\sigma$-model (and also by
the diagrammatic perturbation theory) are in agreement.
In the insulating regime, the behavior of the probability distribution
following from the DMPK-equation is rather well understood.
The Lyapunov exponents are known analytically and the $\lambda_i$
(and hence the conductance $g=\sum_i (1+\lambda_i)^{-1})$)
follow a log-normal distribution
\cite{pichardzanon,lesarcs,review_matrix}.
However, a direct comparison with the
results of the $\sigma$-model is rather difficult because
only the averages $\langle g\rangle$ and $\langle g^2\rangle$
(instead of $\langle \ln g\rangle$) are
known. The issue is rather important due to the unusual results
of the $\sigma$-model for the symplectic symmetry class (systems with
time-reversal symmetry and rather strong spin-orbit coupling):
Both the average conductance and its variance were found to have finite
values as $L/\xi\to\infty$ \cite{zirnbauer1,zirnbauer2}.
It is therefore of considerable interest to derive from the DMPK-equation
closed expressions of the average conductance in the insulating regime,
even though $\langle g\rangle$ does not coincide with the typical
conductance $\exp(\langle \ln g\rangle)$.

The situation in the unitary symmetry class (systems with no time
reversal symmetry) is particularly favorable because
recently \cite{rejaei} the DMPK-equation has been solved
exactly by means of a Sutherland-transformation \cite{sutherland}
leading to an independent Fermion problem.
Based on the expression of $p(\{\lambda_i\},L)$ given in Ref.
\cite{rejaei}, this letter is concerned with the exact calculation
of the $m$-point correlation functions
\begin{eqnarray}
\label{mcorr_def}
&& R_m(\lambda_1,\ldots,\lambda_m,L)  =  \\
\nonumber
&& \quad=\frac{N!}{(N-m)!}
\int_0^\infty d\lambda_{m+1}\cdots d\lambda_N\ p(\{\lambda_i\},L)
\end{eqnarray}
that are usually considered in the theory of random matrices
\cite{mehta}.

The exact result for the joint probability distribution
given in Ref. \cite{rejaei} can be expressed in terms of
the variables $\lambda_i$ as
\begin{equation}
\label{probmello}
p(\{\lambda_i\},t)\propto \prod_{i>j} (\lambda_i-\lambda_j)
\,\det(g_{m-1}(\lambda_j,t))\,e^{C_N t}\quad,
\end{equation}
with
\begin{eqnarray}
\nonumber
g_m(\lambda,t) & = &
\int_0^\infty dk\,{\textstyle (k^2)^m\ \frac{k}{2}\tanh\left(
\frac{\pi k}{2}\right)}\,\\
\label{mellogndef}
&&\times P_{\frac{1}{2}(ik-1)}(1+2\lambda)
e^{-k^2 t}\quad,
\end{eqnarray}
and a constant $C_N$ to be found in \cite{rejaei,frahm1}.
The functions $P_{\frac{1}{2}(ik-1)}(1+2\lambda)$ denote the
generalized Legendre functions.
We have introduced the abbreviation $t=L/2\xi$,
i.e. the system length measured in units of (twice) the localization length.

In random matrix theory, one typically applies
the method of orthogonal polynomials \cite{mehta} to calculate
the $m$-point functions (\ref{mcorr_def}). This method cannot directly
be carried over to the present case
because the pairwise ``interaction'' between the $\lambda_i$
eigenvalues is no longer purely logarithmic \cite{rejaei}.
A first generalization \cite{mutt1} of the method is based on
the application of biorthogonal polynomials. Extending the idea of
Ref. \cite{mutt1}, we write the probability distribution
(\ref{probmello}) as a product of two determinants
\begin{equation}
\label{probmello2}
p(\{\lambda_i\},t)\propto
\det(Q_{n-1}(\lambda_j,t))\,
\det(h_{m-1}(\lambda_j,t))
\end{equation}
with ``dual'' functions $Q_n(\lambda,t)$ and $h_m(\lambda,t)$
($n,m=0,1,\ldots,N-1$) that fulfill the biorthogonality relation
\begin{equation}
\label{quasi_orth}
\int_0^\infty d\lambda\ Q_n(\lambda,t)
\ h_m(\lambda,t)=\delta_{nm}\quad.
\end{equation}
Expressing the product of the $\lambda_i$-differences in (\ref{probmello})
as a Vandermonde determinant with (arbitrary) polynomials
$Q_n(\lambda,t)$ of degree $n$, we arrive at (\ref{probmello2}). The
biorthogonality (\ref{quasi_orth}) can be achieved by a suitable
linear transformation of the functions $g_m(\lambda,t)$.
In the following, we choose the $Q_n(\lambda,t)$ as
\begin{equation}
\label{melloqndef}
Q_{n}(\lambda,t)=P_n(1+2\lambda)\,e^{-\varepsilon_n t}\quad,\quad
\varepsilon_n=(1+2n)^2
\end{equation}
where $P_n$ are the Legendre polynomials.

The next step is the evaluation of the integral of each combination
of $Q_n(\lambda,t)$ with $g_m(\lambda,t)$.
The Legendre functions $P_{\frac{1}{2}(ik-1)}(1+2\lambda)$ (see Eq.
(\ref{mellogndef})) are eigenfunctions of the differential operator
\begin{equation}
\label{diffdef}
D=-\left(4\frac{\partial}{\partial \lambda}\,\lambda(1+\lambda)
\frac{\partial}{\partial \lambda}+1\right)
\end{equation}
with eigenvalue $\,k^2\,$. The operator $D$ is related to
the one particle Hamiltonian ${\cal H}_0$ used in Ref. \cite{rejaei} by a
suitable transformation. In the following, we need three properties of
$g_m(\lambda,t)$, namely
\begin{eqnarray}
\label{gncalc}
g_m(\lambda,t) & = & D^m\,g_0(\lambda,t)\quad,\\
\label{g0calc}
\frac{\partial}{\partial t}\,g_0(\lambda,t) & = & -D\,g_0(\lambda,t)\quad,\\
\label{g0delta}
g_0(\lambda,0) & = & \delta(\lambda)\quad.
\end{eqnarray}
Eqs. (\ref{gncalc}), (\ref{g0calc}) follow directly from (\ref{mellogndef}).
Eq. (\ref{g0delta}) reflects the initial condition for the one particle
Green's function \cite{rejaei}. In order to proceed, we need the
identity
\begin{equation}
\label{int1}
\int_0^\infty d\lambda\ Q_n(\lambda,t)\,g_m(\lambda,t)=
(-\varepsilon_n)^m\quad,
\end{equation}
which follows from the fact that
the Legendre polynomials $P_n(1+2\lambda)$ are also eigenfunctions
of the differential operator $D$ (with eigenvalue $-\varepsilon_n$).
One can verify for $m=0$ via (\ref{g0calc},\ref{g0delta}) that the
integral in (\ref{int1}) does not depend on $t$ and is equal to unity.
The generalization to arbitrary $m$ follows directly from Eq.
(\ref{gncalc}).

We now construct the functions $h_m(\lambda,t)$ in analogy to Eq.
(\ref{mellogndef}), i.e.
\begin{eqnarray}
\nonumber
h_m(\lambda,t) & = &
\int_0^\infty dk\,{\textstyle L_m(k^2)\ \frac{k}{2}\tanh\left(
\frac{\pi k}{2}\right)}\,\\
\label{mellohndef}
&&\times P_{\frac{1}{2}(ik-1)}(1+2\lambda)
e^{-k^2 t}\quad.
\end{eqnarray}
Here, the $L_m(\cdots)$ denote a set of linearly independent polynomials of
(maximal) degree $N-1$. The integral (\ref{int1}) with the $g_m$ replaced
by $h_m$ takes the value $L_m(-\varepsilon_n)$. The biorthogonality
relation (\ref{quasi_orth}) is therefore fulfilled if we choose
the polynomials $L_m$ as the Lagrangian interpolation
polynomials
\begin{equation}
\label{lagrange_def}
L_m(z)=\prod_{l=0,(l\neq m)}^{N-1}\frac{z-(-\varepsilon_l)}
{(-\varepsilon_m)-(-\varepsilon_l)}\quad.
\end{equation}

As a key step \cite{mutt1} to apply standard methods from random
matrix theory \cite{mehta}, we consider the function
\begin{equation}
\label{mellokndef}
K_N(\lambda,\tilde\lambda;t)=\sum_{m=0}^{N-1}
Q_m(\lambda,t)\,h_m(\tilde\lambda,t)\quad
\end{equation}
with the two important properties
\begin{eqnarray}
\label{melloknnorm}
\int_0^\infty d\lambda\ K_N(\lambda,\lambda;t)&=&N\quad,\\
\label{melloknkonv}
\int_0^\infty d\mu\ K_N(\lambda,\mu;t)\,K_N(\mu,\tilde\lambda;t)&=&
K_N(\lambda,\tilde\lambda;t)\quad.
\end{eqnarray}
The application of a general theorem of the theory of random matrices
\cite{dyson4,discuss} immediately yields the $m$-point correlation functions
\begin{equation}
\label{mellocorrfunc}
R_m(\lambda_1,\ldots,\lambda_m;t)=\det
\left(K_N(\lambda_i,\lambda_j;t)_{\,1\le i,j\le m}\right)\quad.
\end{equation}
This expression together with Eqs.
(\ref{melloqndef},\ref{mellohndef},\ref{mellokndef}) forms
the key result of this letter. For $m=1$
($m=2$) it represents the density (the two-point function)
\begin{eqnarray}
\label{dens1}
R_1(\lambda;t) & = & K_N(\lambda,\lambda;t)\quad,\\
\label{dens2}
R_2(\lambda,\tilde\lambda;t) & = & R_1(\lambda;t)\,R_1(\tilde\lambda;t)\\
\nonumber
&& -K_N(\lambda,\tilde\lambda;t)\,K_N(\tilde\lambda,\lambda;t)\quad.
\end{eqnarray}
In the insulating regime $t\gg 1$ the $k$-integration
in (\ref{mellohndef}) can be evaluated by a saddle-point
approximation \cite{frahm1} which yields for
the density of the variable $x=\mbox{arsinh}(\sqrt\lambda)$
a sum of Gaussian distributions with
centers $2t(1+2 m)$ ($m=0,1,\ldots,N-1$) and variances $2t$.
This result reflects the self averaging
behavior of $x$ in the localized regime and is
well understood by a direct analysis of the DMPK-equation
\cite{pichardzanon,lesarcs,review_matrix} in terms of
Lyapunov exponents.
The results of a numerical evaluation of (\ref{mellohndef})
for $N=5$ and three typical length scales ($L/\xi=0.2,\ 1,\ 10$)
which correspond to the metallic, crossover, and insulating regime
are given in Fig. \ref{fig1} which shows the density
$\rho(x)=\sinh(2x)\,R_1(\sinh^2 x,\,L/(2\xi))$ versus $x$.
Both axis have been rescaled with the factor $L/\xi$ for convenience.
In all three cases, the effect of the level repulsion
is easily recognized because it leads to distinct maxima and
minima in the level density. This effect is most prominent in the
localized regime (curve (a), $L=10\,\xi$) where each level
has his own Gaussian peak well separated from the
other levels. In the crossover regime (curve (b), $L=\xi$)
the levels begin to mix but the amplitudes of the
maxima are still rather large in comparison to the minima. Even in the
so-called metallic regime (curve (c), $L=0.2\,\xi$) the oscillations
are still visible and the density is not entirely constant.
On the other hand, a constant density is expected for this regime
\cite{rejaei}.
This discrepancy is a finite $N$ effect (note that $N=5$ in Fig.
\ref{fig1}). The usual metallic regime, $l\ll L \ll \xi=2Nl$,
exists in the large $N$ limit only.

Now, we focus attention on the average conductance and its second
moment. The conductance is expressed in terms of the density
(\ref{dens1}) as
\begin{equation}
\label{melloavcond}
\langle g\rangle =\int_0^\infty d\lambda\,\frac{1}{1+\lambda}
\sum_{m=0}^{N-1} Q_m(\lambda,t)\,h_m(\lambda,t)\quad.
\end{equation}
Since
$Q_m(\lambda,t)$ is a polynomial of degree $m$ in the variable
$1+\lambda$, we can use the decomposition
\begin{equation}
\label{qmpol}
\frac{1}{1+\lambda}Q_m(\lambda,t)
=\frac{(-1)^m}{1+\lambda}\,e^{-\varepsilon_m t}+
r_{m-1}(\lambda)
\end{equation}
where $r_{m-1}(\lambda)$ is a polynomial of degree $m-1$ that does not
contribute in the integral (\ref{melloavcond}) due to the
biorthogonality between the $Q_n$ and the $h_m$.
The $\lambda$-integration can be done \cite{frahm1}
resulting in
\begin{eqnarray}
\nonumber
\langle g\rangle & = & 2 \sum_{m=0}^{N-1} \int_0^\infty dk\,
e^{-((1+2m)^2+k^2) t}\ {\textstyle
k\tanh\left(\frac{\pi k}{2}\right)}\\
\label{melloavcond3}
&&\times\frac{2m+1}{k^2+(1+2m)^2}
\ a(N,m,k)\quad,
\end{eqnarray}
where we have introduced the coefficient
\begin{equation}
\label{anmkresult}
a(N,m,k) = \frac{\Gamma\left(N+\frac{1}{2}+i\frac{k}{2}\right)
\ \Gamma\left(N+\frac{1}{2}-i\frac{k}{2}\right)}{
\Gamma(N-m)\ \Gamma(N+m+1)}\quad.
\end{equation}
Using the expressions (\ref{dens1}) and (\ref{dens2}) for
the density and the two-point function, it is also possible to get
by a straightforward calculation the second moment of the conductance
\begin{eqnarray}
\nonumber
\langle g^2\rangle & = &
\frac{1}{2} \sum_{m=0}^{N-1} \int_0^\infty dk\,
e^{-((1+2m)^2+k^2) t}\ {\textstyle
k\tanh\left(\frac{\pi k}{2}\right)}\\
\label{mellofluct}
&&\times(2m+1)\ a(N,m,k)\quad.
\end{eqnarray}
We note that this result can also be derived in
a more direct way by the general identity
\begin{equation}
\label{gderiv_beta}
\frac{\partial \langle g\rangle}{\partial t}=
-4 \left(\langle g_2\rangle + \frac{\beta}{2}\left(
\langle g^2\rangle-\langle g_2\rangle\right)\right)
\end{equation}
that follows \cite{mello4} directly from the original Fokker-Planck equation
for arbitrary $\beta$ (with
$g_2=\sum_i (1+\lambda_i)^{-2}$). In the unitary case $\beta=2$
the average $\langle g_2 \rangle$ drops out and
$\langle g^2\rangle$ is entirely determined by
(\ref{melloavcond3}) and (\ref{gderiv_beta}).

Eqs. (\ref{melloavcond3}) and (\ref{mellofluct}) are
exact for all values of $N$ and $t=L/(2\xi)$.
In the insulating regime, they can be simplified to give
\begin{equation}
\label{gg_insul}
\langle g\rangle\simeq 4\,\langle g^2\rangle\simeq
\frac{\pi^{3/2}}{4}\,a(N,0,0)\,\Bigl(\frac{L}{2\xi}\Bigr)^{-3/2}\,e^{-L/2\xi}
\quad,
\end{equation}
where the coefficient $a(N,0,0)$ takes values between $\pi/4$ for $N=1$
and $1$ for $N\to\infty$. Eq. (\ref{gg_insul}) differs considerably
from the behavior of the typical conductance:
$\exp(\langle \ln g\rangle)\sim e^{-2L/\xi}$.

We can also consider the limit $N\to \infty$
for all regimes. Then, the coefficient
$a(N,m,k)$ tends to unity and the range of the $m$-sum is
extended to infinity. In this case, the resulting expressions
for the average conductance and the second moment become identical with
those obtained
from the one-dimensional non-linear $\sigma$-model (for the unitary case) by
Zirnbauer et al. \cite{zirnbauer1,zirnbauer2}.
(There is still a factor $2$ that accounts for a different consideration
of the spin degeneracy in the Landauer formula.)

This exact agreement for the whole range of length scales from
the metallic to insulating regime strongly suggests that the
Fokker-Planck approach \cite{dorokhov,mello2}
which leads to the DMPK-equation
is completely equivalent (in the large $N$-limit)
to the quasi one-dimensional microscopic
models from which the one-dimensional $\sigma$-model
is derived \cite{efetov,iida,fyodorov}. We have proven this equivalence
for the first two conductance moments.

This result sheds new light on
the symplectic case at $\beta=4$ where a similar agreement seems
hard to believe. At the moment, exact expressions of
$\langle g\rangle$ and $\langle g^2\rangle$
for the DMPK-equation and $\beta=4$ are not
known but it is nevertheless clear that the typical conductance
decreases exponentially with the length of the wire. In fact,
the Lyapunov exponents are known for every value
of $\beta=1,2,4$ \cite{pichardzanon,lesarcs,review_matrix} and
one expects for the density of the $x$-variable in the
insulating regime a similar behavior as in Fig. \ref{fig1} (a), with Gaussian
maxima at the positions $2t(1+\beta m)$. The typical conductance
corresponds to the first maximum whereas the average conductance
$\langle g\rangle =\int dx\ \rho(x)\,\cosh^{-2}(x)$
is determined by the density at small $x$ where
the deviations from the Gaussian approximation are rather strong. A
precise quantitative estimation of
$\langle g\rangle$ is thus not possible within this simple
picture. In Ref. \cite{hueffmann} it was found by an involved and
approximate calculation that the average conductance decreases
exponentially with the length of the wire for all values of $\beta$.

On the other hand, a simple and precise argument
can be given by the exact expression (\ref{gderiv_beta}). Let us
assume that the average conductance
approaches exponentially its ``finite non-vanishing limit'' for
$t\to\infty$, as suggested by the $\sigma$-model result for $\beta=4$.
Then $\partial\langle g\rangle/\partial t$
vanishes and we get from Eq. (\ref{gderiv_beta}) and the
inequality $0\le g_2\le g^2$ immediately
$\langle g_2\rangle=\langle g^2\rangle=0$ thus contradicting the
assumption.

Apparently, the Fokker-Planck approach
and the supersymmetric description for quasi one-dimensional
conductors seem to be equivalent for $\beta=2$ and disagree
in the insulating regime for $\beta=4$. It must be
emphasized that the unusual results of the $\sigma$-model
approach \cite{zirnbauer1,zirnbauer2} are due to a so-called
``zero mode contribution'' which is absolutely necessary \cite{zirnbauer2}
to reproduce the correct behavior in the metallic regime.
At the moment, it is not clear whether there is
a particular problem with the $\sigma$-model, probably related to
the large $N$ limit, or whether in the Fokker-Planck approach
the assumption of an {\em isotropic\/}
diffusion in transfer matrix space needs revision.
This assumption could in principle be wrong just for $\beta=4$.
A satisfactory answer to these questions remains an outstanding
and important problem.

The author thanks J.-L. Pichard, C. W. J. Beenakker, and
M. R. Zirnbauer for helpful discussions and A. M\"uller-Groeling for
reading the manuscript. The D.F.G. is acknowledged for a
post-doctoral fellowship.

\begin{figure}
\caption{The density $\rho(x)\,L/\xi$ of the variable
$x=$ $\mbox{arsinh}(\protect\sqrt\lambda)$ versus $x\,\xi/L$
for the cases $L=10\,\xi$ (a), $L=\xi$ (b), and $L=0.2\,\xi$ (c).}
\label{fig1}
\end{figure}

\includegraphics{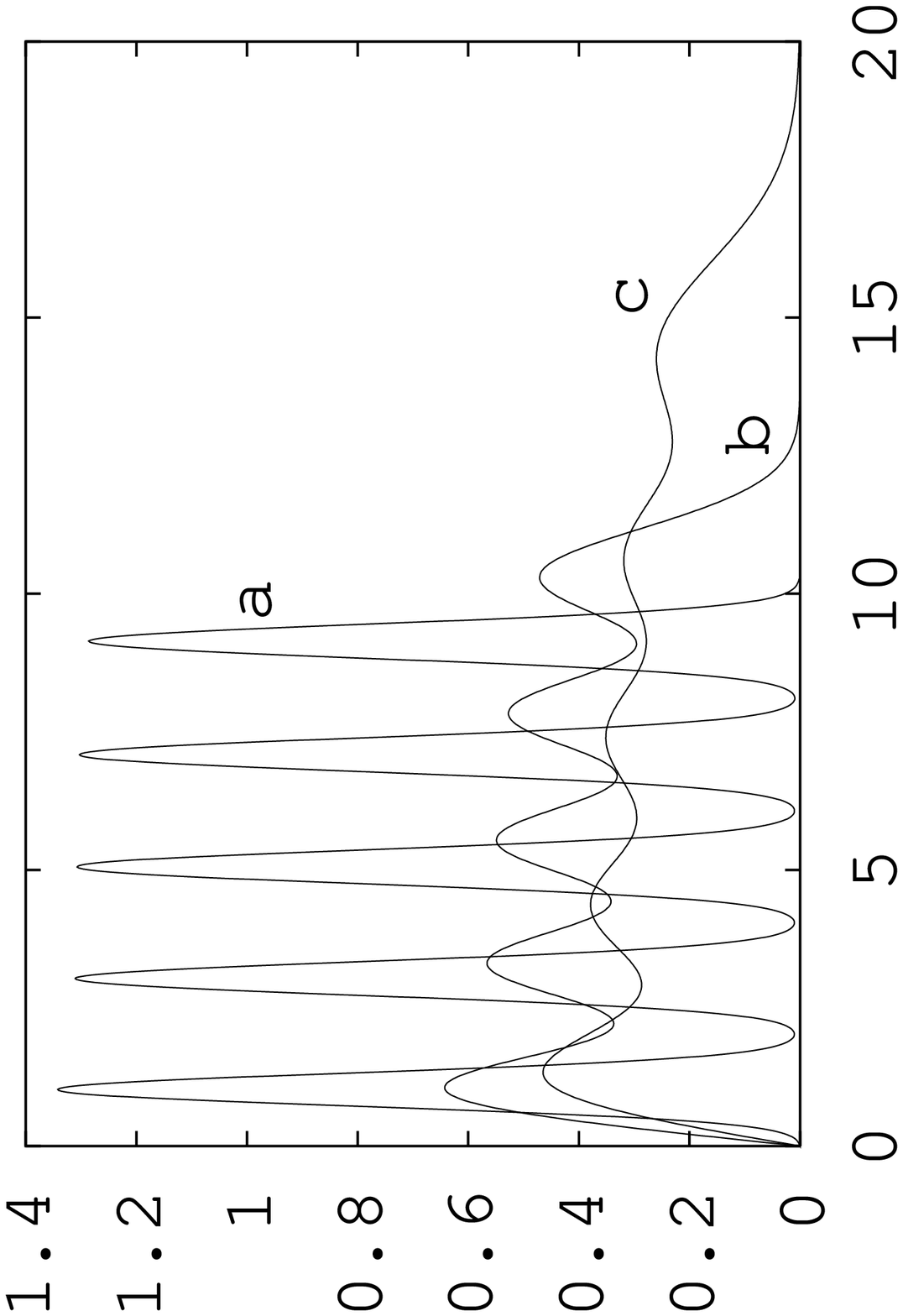}
\centerline{\LARGE Fig. 1}
\vfill\eject

\end{document}